\documentclass[aps,prl,twocolumn,floats]{revtex4}

\usepackage{amssymb}
\usepackage{amsmath}
\usepackage{graphicx}
\usepackage{bm}
\usepackage{mdframed}
\usepackage{color}
\usepackage{gensymb}
\usepackage{comment}

\begin{document}

\bibliographystyle{prsty}

\title{Spin-transfer and Topological Hall Effects as Novel Probes for Magnetic Disclinations in Frustrated Magnets}

\author{Ricardo Zarzuela$^{1}$ and Jairo Sinova$^{1,2}$}

\affiliation{$^{1}$Institut f\"{u}r Physik, Johannes Gutenberg Universit\"{a}t Mainz, D-55099 Mainz, Germany \\$^{2}$Institute of Physics Academy of Sciences of the Czech Republic, Cukrovarnick\'{a} 10, 162 00 Praha 6, Czech Republic
}

\begin{abstract}
Magnetic frustrated systems have resurged in spintronics as optimal candidates for hosting three dimensional topological solitons, such as Shankar skyrmions and 4$\pi$-vortices, and other singular topological defects. These topological excitations are encoded in the order-parameter connected to the spin-spin correlation of the system. The key challenge for their experimental discovery is the lack of probes to observe them. We demonstrate here that spin-transfer torque and topological Hall effect measurements can serve as probes of magnetic disclinations and solitons in these systems, by means of a previously unidentified contribution to them from these topological defects, with no analog in collinear magnetism. We present a minimal low-energy long-wavelength theory for the itinerant carriers and derive the effective emergent electrodynamics arising from the noncoplanar magnetic background with topological defects and solitons. This opens new avenues for the detection of topological defects in magnetic systems with order-parameter manifolds beyond the conventional $S^{2}$ (unit sphere) ferromagnetic paradigm.
\end{abstract}
\maketitle

\section{Introduction}
Topological magnetic soliton excitations, such as skyrmions in ferromagnetic systems, have emerged as novel quasiparticles whose manipulation can be exploited for future spintronic memory and computing devices. At the same time, topological defects, which represent singularities in the order-parameter manifold, are ubiquitous in physics and originate in the spontaneous breaking of the underlying symmetry \cite{Kibble-JPA1976,Kibble-PR1980,Zurek-Nature1985,Zurek-APPB1993,Zurek-PR1996,delCampo-IJMPA2014}, e.g., point defects, dislocations and disclinations in elastic media \cite{Romanov-PSSA1983,Kleman-RMP2008} and liquid crystals \cite{Lavrentovich-Springer2001,Andrienko-JML2018}. Skyrmions in collinear magnetic systems exist within the $S^2$ (unit sphere) manifold, which has a limited number of topological excitations. More complex topological magnetic solitons, such as Shankar skyrmions \cite{Shankar-JPhys1977,Volovik-JETP1977} and Anderson-Toulouse $4\pi$-vortices \cite{Anderson-PRL1976}, which are intrinsically three-dimensional, are ruled out in this order-parameter manifold by homotopy arguments \cite{Mermin-RMP1979}. The most natural magnetic platforms to be able to host these exotic solitons are magnetic frustrated systems dominated by isotropic exchange interactions, here referred to as \textit{frustrated magnets}. Geometric frustration in these systems manifests itself through a disclination network in the spin sector \cite{Toulouse-CommPhys1977,Villain-JPhys1977,Toulouse-PhysRep1979}. These platforms exhibit no long-range magnetic order at mesoscopic length scales and are described by a spin-spin correlator, which yields an order-parameter manifold consisting of rotation matrices (namely, the SO(3) group) \cite{Henley-AnnPhys1984a,Edwards-JPF1975}.

Based on a recent theory for transport of itinerant carriers in a magnetically frustrated conductor, which appears in a companion paper \cite{Zarzuela-PRB2021}, in this Letter we demonstrate the possibility of utilizing the spin-transfer \cite{Ralph-JMMM2008,Gomonay-LTP2014,Baltz-RMP2018} and the topological Hall effects \cite{Taguchi-Science2001,Yanagihara-PRL2002,Machida-PRL2007,Neubauer-PRL2009,Kanazawa-PRL2011,Yu-NatMater2011,Ueda-PRL2012,Huang-PRL2012,Shiomi-PRB2012,Li-PRL2013,Surgers-NatComms2014,Leroux-SciRep2018,Kurumaji-Science2019,Shao-NatElectron2019,Yu-JMMM2019,Zhang-APL2019,Li-APL2019,He-APL2020,Lim-Small2020,Ghimire-SciAdv2020,Wang-PRB2021} as sensitive probes for magnetic disclinations and unconventional SO(3)-solitons emerging in frustrated metallic magnets. These effects have been successful in detecting nontrivial spin textures in ferromagnets, but were previously expected to not exist in frustrated ferromagnets due to their zero magnetization at the mesoscopic scales. Within a semiclassical approach, we derive two new emergent main contributions to the spin-transfer torque and the Hall current in these platforms: the first one depends on an effective nonabelian magnetic field, generated by the aforesaid topological solitons, whereas the second contribution is parametrized by the density of magnetic disclinations. Remarkably, the obtained spin-transfer torque finds no counterpart in collinear magnetism. In the Hall scenario, however, only the first contribution can be identified with the topological Hall current observed in collinear magnets. Therefore, these previously unexplored (topological) defect-mediated spintronic effects can have important consequences beyond our model and represent one of the smoking guns of topological transport in frustrated spin systems.

\begin{figure*}[ht!]
\begin{center}
\includegraphics[width=1.0\textwidth]{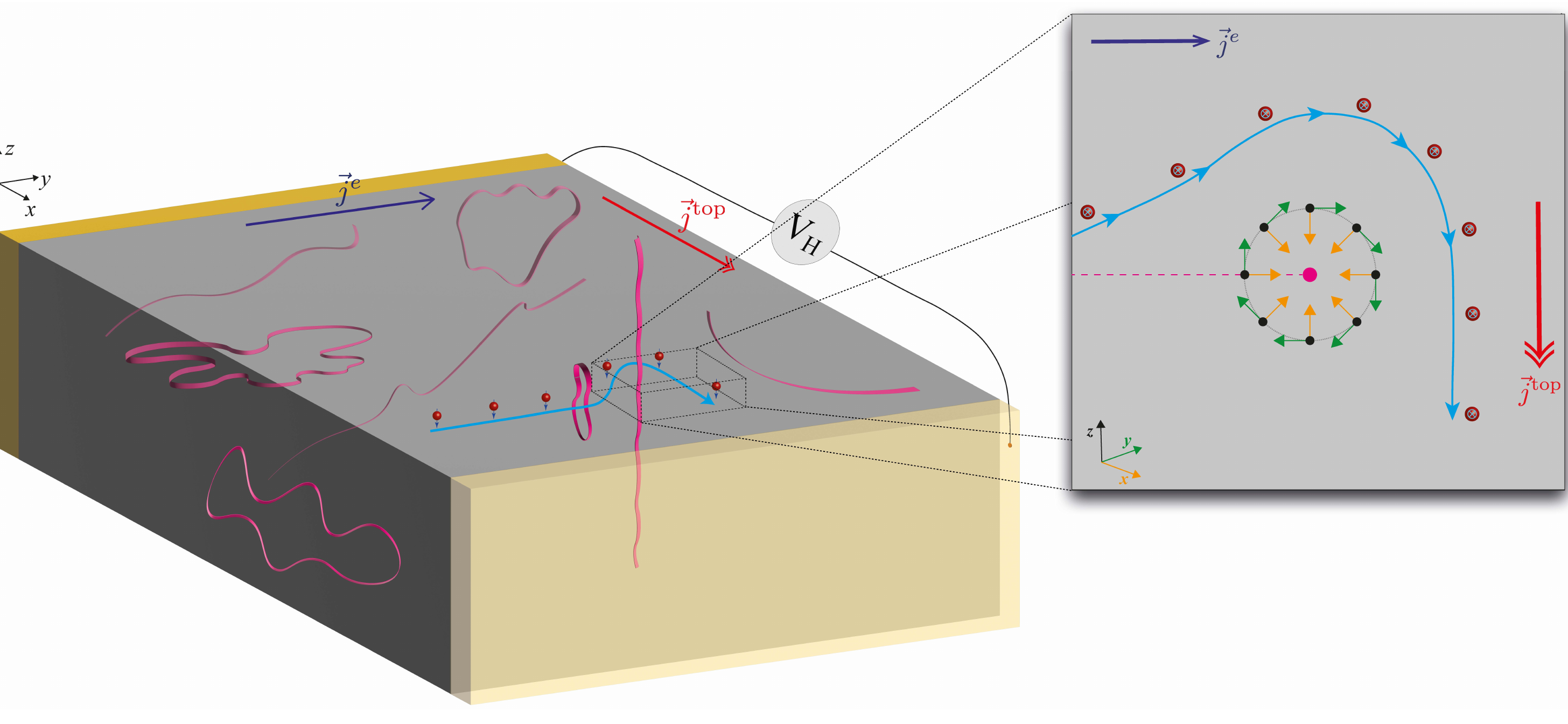}
\caption{A schematic of the magnetotransport device considered for probing the disclination-mediated topological Hall effect in frustrated platforms. A rectangular thin film made of a frustrated magnet (grey) is sandwiched between two metallic contacts (gold). A charge current ($\vec{j}^{e}$, blue arow) is injected and a voltage is measured in a Hall geometry. The charge carriers (dark red spheres crossed by a blue arrow) are deflected (blue path) due to the presence of an effective (real-space) magnetic field, proportional to the projection of the itinerant spin polarization onto the magnetic disclination density vector. The resulting Hall current $\vec{j}^{\textrm{top}}$ (red arrow) can be detected as a Hall voltage $V_{H}$. (Inset) Hall-type trajectory of the itinerant carriers induced by a quasi-straight magnetic disclination extending along the $z$ direction. The projection of the magnetic disclination onto a cross section ($xy$ plane) of the sample resembles a magnetization vortex from collinear magnetism: the spin frame of reference, denoted by a triad of orange ($x$ axis), green ($y$ axis) and black ($z$ axis) vectors, rotates an angle $2\pi$ (for a disclination of strength $\Omega/2\pi=+1$) with respect to the $z$ axis when revolving around the disclination core (magenta dot). This emphasizes the multi-valued nature of the SO(3)-order parameter across the essential branch cut (magenta dashed line) of this topological singularity.}
\vspace{-0.5cm} 
\label{Fig1}
\end{center}
\end{figure*}

\textit{Effective model and disclination network.}|The effective low-energy long-wavelength Hamiltonian describing the physics of itinerant carriers in a magnetically frustrated conductor is given by \cite{Zarzuela-PRB2021}:
\begin{equation}
\label{eq:eff_Ham1}
\hat{H}_{\textrm{eff}}=\tfrac{1}{2m_{\star}}\big(\vec{p}- \hbar g\,\vec{\bm{\Omega}}\circ\bm{\tau}\big)^{2}+\tfrac{\hbar}{2}\bm{\omega}_{m}\circ
\bm{\tau}-\Phi_{m}\tau_{0},
\end{equation}
where $\vec{p}=-i\hbar\vec{\nabla}$ is the momentum operator, $\tfrac{\hbar}{2}\bm{\omega}_{m}=g_{0}\bm{\Omega}_{t}-g_{4}\bm{m}$ and $\Phi_{m}=g_{2}\bm{m}^{2}+g_{3}\sum_{\mu=x,y,z}\bm{\Omega}_{\mu}\circ\bm{\Omega}_{\mu}$ define an effective spin precession vector and electric potential for the itinerant carriers, respectively. Here $\tau_{0}$ is the identity, $\bm{\tau}$ denotes the vector of Pauli matrices,  $g$ and $\{g_{k}\}_{k=0,2,3,4}$ are coupling constants, and $m_{\star}$ is the effective carrier mass.  In our notation, bold and $\vec{\,}\,$ denote vectors in  the spin and real spaces, respectively. Furthermore, $\circ$ denotes the dot product in spin space, $\{\bm{\Omega}_{\mu}\}_{\mu=t,x,y,z}$ denote the Yang-Mills tensor fields specifying the local curved spin geometry of the frustrated magnet \cite{Dzyaloshinskii-AnnPhys1980} and $\bm{m}$ is its macroscopic spin density. 

 The Heisenberg equation $\frac{d \vec{x}}{d t}=\frac{1}{i\hbar}\big[\vec{x},\hat{H}_{\textrm{eff}}\big]$ for the position operators yields the expression $\vec{\Pi}\equiv m_{\star}\frac{d \vec{x}}{dt}=$  $\vec{p}-\hbar g\,\vec{\bm{\Omega}}\circ\bm{\tau}$ for the kinetic momentum; thereby Eq.~\eqref{eq:eff_Ham1} can be recast as
\begin{equation}
\label{eq:eff_Ham2}
\hat{H}_{\textrm{eff}}=\tfrac{1}{2m_{\star}}\vec{\Pi}^{2}+\tfrac{\hbar}{2}\bm{\omega}_{m}\circ\bm{\tau}-\Phi_{m}\tau_{0}.
\end{equation}

A hydrodynamic theory for the disclination network encoded within the Yang-Mills tensor fields, starts from the disclination density tensor, $\hat{\bm{\rho}}$ and the disclination currents, $\vec{\bm{\mathcal{J}}}$, which are given locally by \cite{Volovik-ZhETF1978,Dzyaloshinskii-AnnPhys1980}:
\begin{align}
\label{discl_dens}
\bm{\rho}_{\mu\nu}&=\partial_{\mu}\bm{\Omega}_{\nu}-\partial_{\nu}\bm{\Omega}_{\mu}-\bm{\Omega}_{\mu}\bm{\otimes}\bm{\Omega}_{\nu},\\
\label{discl_curr}
\bm{\mathcal{J}}_{\hspace{-0.05cm}\mu}&=\partial_{t}\bm{\Omega}_{\mu}-\partial_{\mu}\bm{\Omega}_{t}-\bm{\Omega}_{t}\bm{\otimes}\bm{\Omega}_{\mu},
\end{align}
in terms of the Yang-Mills fields, with $\mu,\nu=x,y,z$ running over spatial indices and $\otimes$ being the cross product in spin space. These expressions embody the fact that the disclination network gives rise to the curvature of the Yang-Mills connection arising in the spin space geometry of a frustrated magnet \cite{Footnote00}. Furthermore, due to its antisymmetric nature in real space indices, we can parametrize the tensor~\eqref{discl_dens} in terms of the disclination density axial vector $\vec{\bm{\varrho}}\equiv\tfrac{1}{2}\epsilon_{\mu\nu\sigma}\bm{\rho}_{\mu\nu}\hat{e}_{\sigma}$.

\textit{Emergent electrodynamics.}|The chiral spin background engenders an effective (nonabelian) electromagnetic field affecting the dynamics of the itinerant carriers, whose Faraday tensor is given by $\bm{f}_{\mu\nu}=\partial_{\mu}\bm{\Omega}_{\nu}-\partial_{\nu}\bm{\Omega}_{\mu}+2g\,\bm{\Omega}_{\mu}\bm{\otimes}\bm{\Omega}_{\nu}$ \cite{Yang-PR1954,FN1}. Using Eqs.~\eqref{discl_dens} and~\eqref{discl_curr}, its components can be recast as
\begin{align}
\label{Farad_tensor}
\bm{f}_{\mu\nu}&=\left(1+2g\right)\bm{\Omega}_{\mu}\bm{\otimes}\bm{\Omega}_{\nu}+\bm{\rho}_{\mu\nu},\\
\bm{f}_{0\nu}&=\left(1+2g\right)\bm{\Omega}_{t}\bm{\otimes}\bm{\Omega}_{\nu}+\bm{\mathcal{J}}_{\hspace{-0.05cm}\nu},
\end{align}
{\it i.e.}, the emergent electromagnetic response presents both 'smooth' and 'singular' contributions: the former is proportional to the cross product (in spin space) of the Yang-Mills fields, whereas the latter contains all the information concerning the disclination network and is parametrized by the corresponding disclination density and currents. In turn, the effective electric and magnetic fields are given by $\bm{E}_{k}=-\bm{f}_{0k}$ and $\bm{B}_{k}=\frac{1}{2}\epsilon_{kij}\bm{f}_{ij}$, $k=x,y,z$, 
whose smooth and singular parts read
\begin{align}
\label{E_field}
\vec{\bm{E}}^{\textrm{smth}}&=\left(1+2g\right)\vec{\bm{\Omega}}\bm{\otimes}\bm{\Omega}_{t},\hspace{0.5cm}\vec{\bm{E}}^{\textrm{discl}}=-\vec{\bm{\mathcal{J}}},\\
\label{B_field}
\bm{B}^{\textrm{smth}}_{k}&=\left(\tfrac{1}{2}+g\right)\epsilon_{kij}\,\bm{\Omega}_{i}\bm{\otimes}\bm{\Omega}_{j},\hspace{0.5cm}\vec{\bm{B}}^{\textrm{discl}}=\vec{\bm{\varrho}}.
\end{align}
Consistent with the emergent magnetic field behaviour, the commutation relations for the kinetic momentum operators can be cast as  \cite{SM}
\begin{equation}
\label{comm_rel_kin_mom}
\left[\Pi_{i},\Pi_{j}\right]=i\hbar^{2}g\,\epsilon_{kij}\bm{B}_{k}\circ\bm{\tau}.
\end{equation}
\textit{Spin-transfer physics}.|As we have discussed in Ref.~\onlinecite{Zarzuela-PRB2021}, the spin-transfer response of the itinerant charge fluid described by the effective Hamiltonian~\eqref{eq:eff_Ham1} is of the form $\bm{\tau}=\partial_{t}\bm{m}\propto \vec{j}^{e}\cdot\vec{\bm{\Omega}}$ to the lowest order in the injected current $\vec{j}^{e}$ and the Yang-Mills fields. However, in the Dresselhaus-like scenario (namely, when centrosymmetry in the magnet is fully broken), the spin-transfer torque $\bm{\tau}_{D}=\eta_{D}\,\vec{j}^{e}\cdot(\vec{\nabla}\times\vec{\bm{\Omega}})$ is also allowed by symmetry. This torque is dissipative in nature (odd under time-reversal symmetry) and, microscopically, originates from an additional contribution to the aforesaid effective low-energy long-wavelength (Hamiltonian) description of the itinerant carriers \cite{SM}:
\begin{equation}
\label{eq:add_eff_Ham}
\hat{H}_{\textrm{eff}}^{D}=\frac{1}{2m_{\star}'}\big(\vec{p}- \hbar g_{D}\,[\vec{\nabla}\times\vec{\bm{\Omega}}]\circ\bm{\tau}\big)^{2}.
\end{equation}
Here, $g_{D}$ and $m_{\star}'$ denote a new coupling constant and a second effective carrier mass, respectively. We note that this contribution can be interpreted as resulting from a Yang-Mills-type minimal coupling of the itinerant degrees of freedom to the emergent (nonabelian) magnetic field; therefore, it arises in the expansion of the effective Hamiltonian in terms of the derivatives of the Yang-Mills fields, with Eq.~\eqref{eq:eff_Ham1} representing the zeroth-order term. With account of Eqs.~\eqref{discl_dens} and~\eqref{B_field} for the disclination density and the emergent magnetic field, we can recast this spin-transfer torque as
\begin{equation}
\label{st_torque}
\bm{\tau}_{D}=\eta_{D}\,\vec{j}^{e}\cdot{\vec{\bm{\varrho}}}+\frac{\eta_{D}}{1+2g}\vec{j}^{e}\cdot\vec{\bm{B}}^{\textrm{smth}},
\end{equation}
where the prefactor reads $\eta_{D}=\hbar g_{D}e/m_{\star}'\vartheta g_{4}$ in terms of microscopic parameters, and $\vartheta$ is the electrical conductivity of the magnet. Therefore, we conclude that magnetic disclinations mediate an unconventional spin-transfer effect in frustrated magnets, which is one our main findings. The corresponding magnetic torque is linear in both the injected current and the disclination density (vector) and can be tested experimentally: in the absence of external magnetic fields and SO(3) solitons in the spin background, these disclination-induced torques will trigger the magnetization dynamics of the system from an equilbrium static configuration and, in turn, the dynamical generation of a nonequilibrium spin density $\bm{m}$. 
Its measurement would provide a quantitative probe of the density of magnetic disclinations, whose (spatial) components can be extracted by changing the direction of the injected current.

\textit{Hall response.}|The electric current flowing within the magnetic conductor can be estimated via a Drude model for semiclassical transport \cite{Chudnovsky-PRL2007}. The charge current operator, $\vec{j}\equiv e\frac{d\vec{x}}{dt}$, and the steady-state dynamics of the itinerant carriers is described by the equation \cite{SM}:
\begin{widetext}
\begin{align}
\label{EoM_Drude2}
\frac{d\vec{x}}{dt}&=\mu\left[\vec{E}_{\textrm{ext}}-\tfrac{2}{|e|}\left(g_{2}\bm{m}\circ\vec\nabla\bm{m}+g_{3}\bm{\Omega}_{\nu}\circ\vec\nabla\bm{\Omega}_{\nu}\right)\right]\tau_{0}+\frac{\hbar g}{2m_{\star}}\mu\tau\left[\vec{E}_{\textrm{ext}}\times\big(\vec{\bm{B}}\circ\bm{\tau}\big)-\big(\vec{\bm{B}}\circ\bm{\tau}\big)\times\vec{E}_{\textrm{ext}}\right]\\
&\hspace{0.5cm}+\frac{(\chi g_{4}-g_{0})\mu}{|e|}\left[\vec\nabla\bm{\Omega}_{t}+2g\,\vec{\bm{\Omega}}\bm{\otimes}\bm{\Omega}_{t}\right]\circ\bm{\tau},\nonumber
\end{align}
\end{widetext}
where $\mu=e\tau/m_{\star}$ is the carrier mobility, $\vec{E}_{\textrm{ext}}$ denotes the external driving field, $\tau$ represents the scattering time and $e<0$ is the electron charge. The expectation value of the current operator reads $\textrm{Tr}\big[\hat{\rho}\vec{j}\,\big]=e\textrm{Tr}\left[\hat{\rho}\frac{d\vec{x}}{dt}\right]$; the density matrix operator is expanded as $\hat{\rho}=\frac{n}{2}\left[\tau_{0}+\bm{p}\circ\bm{\tau}\right]$ in the basis of Pauli matrices, where $n=\textrm{Tr}[\hat{\rho}]$ and $\bm{p}=\textrm{Tr}[\hat{\rho}\,\bm{\tau}]/n$ denote the total density and the spin polarization of the itinerant carriers, respectively. The resultant expression for the total electric current takes the form $\vec{j}=\vec{j}^{e}+\vec{j}^{\,\textrm{top}}_{\textrm{smth}}+\vec{j}^{\,\textrm{top}}_{\textrm{discl}}$, where
\begin{widetext}
\begin{align}
\label{top_current_smth}
\vec{j}^{\,\textrm{top}}_{\textrm{smth}}&
=-\tfrac{2\sigma_{D}}{|e|}\left[g_{2}\chi^{2}\bm{\Omega}_{t}\circ\partial_{t}\vec{\bm{\Omega}}+g_{3}(\vec{\bm{\Omega}}\cdot\vec\nabla)\circ\vec{\bm{\Omega}}\right]
+\tfrac{\sigma_{D}(g_{0}-\chi g_{4})}{|e|}\bm{p}\circ\left[\partial_{t}\vec{\bm{\Omega}}
+\vec{\bm{E}}^{\textrm{smth}}\right]+\vec{j}^{e}\times\big(\bm{p}\circ\textstyle{\frac{\hbar g\tau}{m_{\star}}}\vec{\bm{B}}^{\textrm{smth}}\big),\\
\label{top_current_discl}
\vec{j}^{\,\textrm{top}}_{\textrm{discl}}&=
\underbrace{\tfrac{2\sigma_{D}}{|e|}\left[g_{2}\chi^{2}\bm{\Omega}_{t}\circ\vec{\bm{\mathcal{J}}}-g_{3}\,\vec{\bm{\Omega}}\circledast\vec{\bm{\varrho}}\right]
\,\,\,\,\,\,\,\,\,\,\,\,\,\,\,\,\,\,\,\,\,\,}_{\text{ effective Coriolis term}}
+\underbrace{\tfrac{\sigma_{D}(\chi g_{4}-g_{0})}{|e|}\bm{p}\circ\vec{\bm{\mathcal{J}}}
\,\,\,\,\,\,\,\,\,\,\,\,\,\,\,\,\,\,\,\,\,\,\,\,\,\,\,\,\,\,\,\,\,}_{\text{effective spin-motive force term}}
\,+\underbrace{\vec{j}^{e}\times\big(\bm{p}\circ\textstyle{\frac{\hbar g\tau}{m_{\star}}}\vec{\bm{\varrho}}\big)\,\,\,\,\,\,\,\,}_{\text{topological Hall term}},
\end{align}
\end{widetext}
are topological contributions to the (Hall) current rooted in the presence of topological solitons and singularities in the SO(3)-order parameter, respectively \cite{SM}. Here, $\vec{j}^{e}=\sigma_{D}\vec{E}_{\textrm{ext}}$ denotes the injected current, $\sigma_{D}=ne\mu$ is the Drude conductivity and $'\circledast'$ stands for the combination of the cross product in real space with the scalar product in spin space, namely $\vec{\bm{A}}\circledast\vec{\bm{B}}=
\vec{\bm{A}}\times\circ\vec{\bm{B}}=\epsilon_{ijk}\hat{e}_{k}A_{i}^{\alpha}B_{j}^{\alpha}$. 

Within these current contributions we identify three types of terms. The first is an effective Coriolis-type term, in the spirit that it arises from the motion of the electron in the background of the rotating local spin geometry specified by the Yang-Mills fields. The second is an effective motive force, arising from the time evolution of the effective smooth electric field originating from the time evolution of the Yang-Mills fields, $\vec{\bm{E}}^{\textrm{smth}}=\left(1+2g\right)\vec{\bm{\Omega}}\bm{\otimes}\bm{\Omega}_{t}$, and the dislocation part $\vec{\bm{E}}^{\textrm{discl}}=-\vec{\bm{\mathcal{J}}}$, from Eq.~\ref{E_field}. The last term is the well known topological Hall term, resultant from the adiabatic motion of the itinerant carrier in the background of the effective magnetic field, which contains again both smooth and disclination-induced parts, as given in Eq.~\ref{B_field}.

Hence, our semiclassical treatment of the spin current operator $\vec{\bm{J}}=\frac{\hbar}{2}\bm{\tau}\frac{d\vec{x}}{dt}$ yields topological contributions to the spin (Hall) current, which are, again, engendered by the presence of solitons and disclinations in the magnetic background. These important contributions in frustrated systems have not been previously identified. A simple set-up for the proposed detection of such a term is shown in Fig.~\ref{Fig1}, where the dislocations that go across the sample thickness can be seeded in the growth process.

\textit{Soliton-mediated effects.}|The SO(3)-order parameter hosts intrinsically three-dimensional solitons, which can induce unconventional spin-transfer and Hall responses in frustrated magnets. In this regard, we focus mainly on the classes of Shankar skyrmions \cite{Shankar-JPhys1977,Volovik-JETP1977}, the condensed-matter realization of skyrmions emerging in low-energy chiral effective descriptions of QCD, and $4\pi$-vortices \cite{Anderson-PRL1976}. The effective magnetic field ascribed to Shankar solitons reads \cite{SM}
\begin{align}
\label{eff_mag_field_spin}
\bm{B}_{k}\propto&\sin\left[f(r)\right]\frac{f'(r)}{r}\left(\hat{\bm{e}}_{k}-\frac{x_{k}}{r}\hat{\bm{e}}_{r}\right)+\frac{4x_{k}}{r^{3}}\sin^{2}\left[f(r)/2\right]\hat{\bm{e}}_{r}\nonumber\\
&+\sin^{2}\left[f(r)/2\right]\frac{f'(r)}{r^{2}}\epsilon_{k\mu\nu}(x_{\nu}\hat{\bm{e}}_{\mu}-x_{\mu}\hat{\bm{e}}_{\nu}),
\end{align}
where $f(r)$ is a smooth radial function that describes the angle of the order-parameter rotation around the radial direction. We note that $f$ satisfies the boundary conditions $f(0)=2\pi$ and $f(\infty)=0$ and that the skyrmion center is located at the origin of the coordinate system. In what follows we assume that the itinerant spin polarization is uniform across the sample and that the topological texture is smooth on the scale of the electron scattering length, i.e. $R_{\star}\ll v_{F}\tau$, where $v_{F}$ and $R_{\star}$ denote the Fermi velocity and the typical size (e.g., radius) of the isolated soliton, respectively. We also assume the uniformity of the spin density $\bm{m}$ across the sample. Thereby, the average  over the soliton volume of both the spin-transfer torque and the topological Hall current is determined by that of the emergent magnetic field. Since its volume-averaged spin-polarized components are $\langle \vec{B}^{x}\rangle=\langle \vec{B}^{y}\rangle=\langle \vec{B}^{z}\rangle=\vec{0}$ \cite{SM}, we conclude that Shankar skyrmions do not mediate either a spin-transfer or a topological Hall response in frustrated magnets.

However, for an isolated $4\pi$-vortex extending along the $z$-axis of the coordinate system we obtain $\langle\vec{B}^{x}\rangle=\langle\vec{B}^{y}\rangle=\vec{0}$ and $\langle\vec{B}^{z}\rangle=\left(1+2g\right)\tfrac{4}{R_{\star}^{2}}\hat{e}_{z}$ \cite{SM}. Therefore, we can conclude that itinerant carriers polarized along the cylindrical axis of the Anderson-Toulouse vortex experience the topological Hall effect mediated by this topological texture, the resultant Hall current reading
\begin{equation}
\langle\vec{j}^{\textrm{\,top}}_{\textrm{$4\pi$}}\rangle=\tfrac{4\hbar\tau}{m_{\star} R_{\star}^{2}}g\left(1+2g\right)(\bm{p}\circ\bm{\hat{e}}_{z})\vec{j}^{e}\times\hat{e}_{z}.
\end{equation}
$4\pi$-vortices also mediate a spin-transfer torque in the direction of its cylindrical axis, which is exerted only by electric currents flowing along the same axis:
\begin{equation}
\langle\bm{\tau}_{D,\textrm{$4\pi$}}\rangle=\tfrac{4\eta_{D}}{R_{\star}^{2}}(\vec{j}^{e}\cdot\hat{e}_{z})\bm{\hat{e}}_{z}.
\end{equation}

O(3) walls constitute another class of solitons that can emerge in frustrated magnets. These topological textures are described by a surface (the so-called wall center, whose dynamics correspond to those of soft modes) across which the order-parameter rotation matrix switches from proper to improper or viceversa \cite{FN2}. O(3) walls are effectively one-dimensional for quasiflat surfaces (e.g., a wall center spanning the $yz$ plane), which leads to a vanishing emergent magnetic field. Therefore, similarly to Shankar skyrmions, they do not mediate in this system neither a spin-transfer nor a topological Hall response.

\textit{Dislocation-mediated effects.}|Conventional imaging techniques do not enable the direct visualization of magnetic disclinations, although scanning nitrogen-vacancy centre magnetometry offers promising perspectives in this regard \cite{Wornle-2019}. However, the spin-transfer and topological Hall effects provide an indirect probe for the presence of $\mathbb{Z}_{2}$-vortex lines/loops in magnetic conductors: to illustrate this, let us consider an isolated disclination line of strength $\Omega/2\pi$ extending along the $z$ axis \cite{FN5}. The corresponding density vector reads $\bm{\rho}_{x}=\bm{\rho}_{y}=\bm{0}$, $\bm{\rho}_{z}=\Omega\delta(x-x_{d})\delta(y-y_{d})\bm{\hat{e}}_{z}$, where $(x_{d},y_{d})$ denotes the centre of the $\mathbb{Z}_{2}$-vortex \cite{SM}. Volume average yields the following expressions for the disclination-mediated spin-transfer torque and charge Hall current:
\begin{align}
\langle\bm{\tau}_{D,\textrm{discl}}\rangle&=\eta_{D}n\Omega(\vec{j}^{e}\cdot\hat{e}_{z})\bm{\hat{e}}_{z},\\
\langle\vec{j}^{\textrm{\,top}}_{\textrm{discl}}\rangle&=\tfrac{\hbar g\tau}{m_{\star}}\eta_{D}n\Omega(\bm{p}\cdot\bm{\hat{e}}_{z})\vec{j}^{e}\times\hat{e}_{z}.
\end{align}
Here $n$ is the areal density of magnetic disclinations and we have assumed that the disclination network is homogeneous over the scale of the electron scattering length, that is, $\ell_{\textrm{discl}}\ll v_{F}\tau$, where $\ell_{\textrm{discl}}$ denotes the typical distance between $\mathbb{Z}_{2}$-vortex lines/loops. Akin to the $4\pi$-vortex scenario, the emergent spin-transfer effect is characterized by the direction of both the spin torque and the electric current coinciding with that of the defect line. Similarly, only carriers polarized along the disclination lines and flowing within the plane perpendicular to it experience the emergent topological Hall response in the frustrated medium. These findings can be tested experimentally via conventional magnetotransport measurements, see Fig.~\ref{Fig1}. In this regard, an estimate of the areal density $n$ can be obtained by pertinent Monte-Carlo simulations of the low-lying spin configurations of the glassy system, along the lines of Ref.~\onlinecite{Henley-AnnPhys1984b}, combined with their analysis via modern visualization techniques for second-rank tensor fields (see, for example, Ref.~\onlinecite{Callan-Jones-PRB2006}, where these techniques are utilized to visualize topological defects in nematic liquid crystals).

\textit{Acknowledgements.}|This work has been supported by the Transregional Collaborative Research Center (SFB/TRR) 173 SPIN+X, the skyrmionics SPP: ZA1194/2-1, SI1720/12-1, Grant Agency of the Czech Republic grant no. 19-28375X, and the Dynamics and Topology Centre funded by the State of Rhineland Palatinate.

\onecolumngrid

\newpage

\hspace{0.1cm}
\section*{\Large Supplemental Material}
\hspace{0.1cm}

\section{Commutation relations for the kinetic momentum}

The commutator of any two components of the kinetic momentum operator can be calculated as follows:
\begin{align}
\label{Comm_rel_kin_mom1}
\left[\Pi_{i},\Pi_{j}\right]&=\big[p_{i}-\hbar g\,\bm{\Omega}_{i}\circ\bm{\tau},p_{j}-\hbar g\,\bm{\Omega}_{j}\circ\bm{\tau}\big]=-\hbar g\big(\underbrace{[p_{i},\bm{\Omega}_{j}\circ\bm{\tau}]}_{-i\hbar\partial_{i}\bm{\Omega}_{j}\circ\bm{\tau}}+\underbrace{[\bm{\Omega}_{i}\circ\bm{\tau},p_{j}]}_{i\hbar\partial_{j}\bm{\Omega}_{i}\circ\bm{\tau}}\big)+\hbar^{2}g^{2}\underbrace{[\bm{\Omega}_{i}\circ\bm{\tau},\bm{\Omega}_{j}\circ\bm{\tau}]}_{2i(\bm{\Omega}_{i}\bm{\times}\bm{\Omega}_{j})\circ\bm{\tau}}\\
&=i\hbar^{2}g\big(\partial_{i}\bm{\Omega}_{j}-\partial_{j}\bm{\Omega}_{i}+2g\,\bm{\Omega}_{i}\bm{\times}\bm{\Omega}_{j}\big)\circ\bm{\tau}=i\hbar^{2}g\,\bm{f}_{ij}\circ\bm{\tau}.\nonumber
\end{align}
Eq.~(9) of the main text stems from the definition of the emergent magnetic field in terms of the Faraday tensor, namely $\vec{\bm{B}}=\frac{1}{2}\epsilon_{kij}\hat{e}_{k}\bm{f}_{ij}$.

\section{Spin-transfer effect mediated by magnetic disclinations}
The derivation of the spin-transfer torque $\bm{\tau}_{D}$ in the Dresselhaus-like scenario follows that of Ref.~20, and exploits the hydrodynamic properties of nonrelativistic Yang-Mills theories [61]: the dynamics of the Fermi field $\Psi$ describing the itinerant carrier are dictated by the total Hamiltonian $\hat{H}_{\textrm{eff}}+\hat{H}^{D}_{\textrm{eff}}$, with equations of motion reading

\begin{align}
\label{EoM1}
\hbar\partial_{0}\Psi^{\dagger}&=-\Phi_{m}\Psi^{\dagger}+\Psi^{\dagger}(\bm{\omega}_{m}\circ\bm{\tau})-\frac{\hbar^{2}}{2\bar{m}_{\star}}\partial_{\mu}^{2}\Psi^{\dagger}-i\frac{\hbar^{2}g}{2m_{\star}}\Psi^{\dagger}(\partial_{\mu}\bm{\Omega}_{\mu}\circ\bm{\tau})-i\frac{\hbar^{2}g}{m_{\star}}\partial_{\mu}\Psi^{\dagger}(\bm{\Omega}_{\mu}\circ\bm{\tau})+\frac{\hbar^{2}g^{2}}{2m_{\star}}\bm{\Omega}_{\mu}^{2}\Psi^{\dagger}\\
&\hspace{2cm} -i\frac{\hbar^{2}g_{D}}{2m_{\star}'}\Psi^{\dagger}(\partial_{\mu}\bm{A}_{\mu}\circ\bm{\tau})-i\frac{\hbar^{2}g_{D}}{m_{\star}'}\partial_{\mu}\Psi^{\dagger}(\bm{A}_{\mu}\circ\bm{\tau})+\frac{\hbar^{2}g_{D}^{2}}{2m_{\star}'}\bm{A}_{\mu}^{2}\Psi^{\dagger},\nonumber\\
\label{EoM2}
\hbar\partial_{0}\Psi&=\Phi_{m}\Psi-(\bm{\omega}_{m}\circ\bm{\tau})\Psi+\frac{\hbar^{2}}{2\bar{m}_{\star}}\partial_{\mu}^{2}\Psi-i\frac{\hbar^{2}g}{2m_{\star}}(\partial_{\mu}\bm{\Omega}_{\mu}\circ\bm{\tau})\Psi-i\frac{\hbar^{2}g}{m_{\star}}(\bm{\Omega}_{\mu}\circ\bm{\tau})\partial_{\mu}\Psi-\frac{\hbar^{2}g^{2}}{2m_{\star}}\bm{\Omega}_{\mu}^{2}\Psi\\
&\hspace{2cm} -i\frac{\hbar^{2}g_{D}}{2m_{\star}'}(\partial_{\mu}\bm{A}_{\mu}\circ\bm{\tau})\Psi-i\frac{\hbar^{2}g_{D}}{m_{\star}'}(\bm{A}_{\mu}\circ\bm{\tau})\partial_{\mu}\Psi-\frac{\hbar^{2}g_{D}^{2}}{2m_{\star}'}\bm{A}_{\mu}^{2}\Psi,\nonumber\\
\end{align}
where $\partial_{0}=-i\partial_{t}$ denotes derivation with respect to the (Wick-rotated) imaginary time, $\bm{A}_{\mu}=\epsilon_{\mu\nu\rho}\partial_{\nu}\bm{\Omega}_{\rho}$ is the $\mu$-th component of the (real-space) curl of the Yang-Mills fields and $\bar{m}_{\star}^{-1}=m_{\star}^{-1}+m_{\star}'^{-1}$. The linear combinations $(\Psi^{\dagger}\tfrac{\hbar}{2}\bm{\tau})$Eq.~\eqref{EoM2}-Eq.~\eqref{EoM1}$(\tfrac{\hbar}{2}\bm{\tau}\Psi)$ and $\Psi^{\dagger}$Eq.~\eqref{EoM2}-Eq.~\eqref{EoM1}$\Psi$ yield the following dynamical equations for the itinerant spin density $\bm{s}=\Psi^{\dagger}\tfrac{\hbar}{2}\bm{\tau}\Psi$ and probability density $\rho=\Psi^{\dagger}\Psi$:
\begin{align}
\label{CE1}
\partial_{t}\bm{s}+\partial_{\mu}\bm{J}_{\mu}&=\bm{\omega}_{m}\bm{\times}\bm{s}+\tfrac{2\bar{m}_{\star}}{\hbar^{2}}\bm{J}_{\mu}\bm{\times}\left(\tfrac{\hbar^{2}g}{m_{\star}}\bm{\Omega}_{\mu}+\tfrac{\hbar^{2}g_{D}}{m_{\star}'}\bm{A}_{\mu}\right),\\
\label{CE2}
\partial_{t}\rho+\partial_{\mu}j_{\mu}&=0,
\end{align}
which are complemented by the following constitutive relations for the spin and charge currents, respectively:
\begin{align}
\label{spin-curr}
\bm{J}_{\mu}&=\frac{\hbar^{2} i}{4\bar{m}_{\star}}\left(\partial_{\mu}\Psi^{\dagger}\bm{\tau}\Psi-\Psi^{\dagger}\bm{\tau}\partial_{\mu}\Psi\right)-\frac{\hbar^{2}g}{2m_{\star}}\rho\,\bm{\Omega}_{\mu}-\frac{\hbar^{2}g_{D}}{2m_{\star}'}\rho\,\bm{A}_{\mu},\\
\label{charge_curr}
j_{\mu}&=\frac{\hbar\, i}{2\bar{m}_{\star}}\left(\partial_{\mu}\Psi^{\dagger}\Psi-\Psi^{\dagger}\partial_{\mu}\Psi\right)-\frac{2g}{m_{\star}}\bm{\Omega}_{\mu}\circ\bm{s}-\frac{2g_{D}}{m_{\star}'}\bm{A}_{\mu}\circ\bm{s}.
\end{align}
As discussed in Ref.~20, the electric current can be cast as $\vec{j}^{e}=e\vec{j}=\vartheta\vec{E}_{\textrm{ext}}+[\hat{L}_{\textrm{qs}}+\hat{L}_{\textrm{qs}}^{D}]\bm{f}_{\bm{m}}+\ldots$, where $\vartheta$ is the conductivity constant (we assume a diagonal structure of the conductivity tensor for the sake of simplicity), $\bm{f}_{\bm{m}}=-\tfrac{\delta\mathcal{E}}{\delta\bm{m}}=\tfrac{2g_{4}}{\hbar}\bm{s}+\ldots$ is the thermodynamic force conjugated to the magnetization of the system, and $\hat{L}_{\textrm{qs}}|_{\mu\mu_{1}}=-\tfrac{\hbar eg}{m_{\star} g_{4}}\Omega_{\mu}^{\mu_{1}}$, $\hat{L}_{\textrm{qs}}^{D}|_{\mu\mu_{1}}=-\tfrac{\hbar eg_{D}}{m_{\star}' g_{4}}A_{\mu}^{\mu_{1}}$ are off-diagonal coefficients of the Onsager matrix. Also, from linear response and reciprocity arguments we know that there is a contribution to the magnetic torque of the form $\partial_{t}\bm{m}=\hat{L}_{\textrm{sq}}^{D}\vec{E}_{\textrm{ext}}$, where the Onsager matrix $\hat{L}_{\textrm{sq}}^{D}$ satisfies the reciprocal relation $\hat{L}_{\textrm{sq}}^{D}=-[\hat{L}_{\textrm{qs}}^{D}]^{\top}$. Therefore, we obtain $\tau_{D}^{\mu}=\tfrac{\hbar eg_{D}}{m_{\star}' g_{4}}A_{\mu_{1}}^{\mu}E_{\mu_{1}}$ as discussed in the main text.

\section{Semiclassical approach for the Hall transport}

In this Section we proceed along the lines of Ref.~55. The time derivative of the kinetic momentum can be easily obtained in the Heisenberg picture:
\begin{align}
\label{Heisenberg_eq_kin_mom}
m_{\star}\frac{d^{2}x_{i}}{dt^{2}}&=\frac{d\Pi_{i}}{dt}=\frac{1}{i\hbar}\big[\Pi_{i},\hat{H}_{\textrm{eff}}\big]=\frac{1}{2im_{\star}\hbar}\big[\Pi_{i},\vec{\Pi}^{2}\big]+\frac{1}{i\hbar}\big[\Pi_{i},\bm{\omega}_{m}\circ\bm{\tau}\big]-\frac{1}{i\hbar}\big[\Pi_{i},\Phi_{m}\tau_{0}\big],\\
&=\frac{\hbar g}{2}\epsilon_{ijk}\left[\frac{dx_{j}}{dt}\big(\bm{B}_{k}\circ\bm{\tau}\big)-\big(\bm{B}_{j}\circ\bm{\tau}\big)\frac{dx_{k}}{dt}\right]+\left[g_{4}\left(\partial_{i}\bm{m}+2g\,\bm{\Omega}_{i}\bm{\times}\bm{m}\right)-g_{0}\left(\partial_{i}\bm{\Omega}_{t}+2g\,\bm{\Omega}_{i}\bm{\times}\bm{\Omega}_{t}\right)\right]\circ\bm{\tau}\nonumber\\
&\hspace{0.5cm}+2\left[g_{2}\bm{m}\circ\partial_{i}\bm{m}+g_{3}\bm{\Omega}_{\mu}\circ\partial_{i}\bm{\Omega}_{\mu}\right]\tau_{0}.\nonumber
\end{align}
With account of the Drude model for dissipation, the equation of motion for itinerant carriers in the steady state becomes
\begin{align}
\label{EoM_Drude}
\frac{m_{\star}}{\tau}\frac{d\vec{x}}{dt}&=-\nabla\Phi+\frac{\hbar g}{2}\left[\frac{d\vec{x}}{dt}\times\big(\vec{\bm{B}}\circ\bm{\tau}\big)-\big(\vec{\bm{B}}\circ\bm{\tau}\big)\times\frac{d\vec{x}}{dt}\right]+\left[g_{4}\left(\nabla\bm{m}+2g\,\vec{\bm{\Omega}}\bm{\times}\bm{m}\right)-g_{0}\left(\nabla\bm{\Omega}_{t}+2g\,\vec{\bm{\Omega}}\bm{\times}\bm{\Omega}_{t}\right)\right]\circ\bm{\tau}\nonumber\\
&\hspace{1cm}+2\left[g_{2}\bm{m}\circ\nabla\bm{m}+g_{3}\bm{\Omega}_{\mu}\circ\nabla\bm{\Omega}_{\mu}\right]\tau_{0},
\end{align}
where $\Phi$ denotes the electrostatic potential associated with the external driving field. Next we split the position operator into the sum of two terms, $\vec{x}=\vec{x}_{0}+\vec{x}_{1}$, so that $\vec{x}_{0}$ is driven by the external electric field (providing therefore the dominant contribution to the total speed) and $\vec{x}_{1}$ is driven by the remaining terms of Eq.~\eqref{EoM_Drude}, which contribute at the subleading order to the speed:
\begin{align}
\label{EoM_Drude2a}
\frac{d\vec{x}_{0}}{dt}&=-\frac{\tau}{m_{\star}}\nabla\Phi=\frac{e\tau}{m_{\star}}\vec{E}_{\textrm{ext}},\\
\label{EoM_Drude2b}
\frac{d\vec{x}_{1}}{dt}&=\frac{\hbar g\,\tau}{2m_{\star}}\left[\frac{d\vec{x}}{dt}\times\big(\vec{\bm{B}}\circ\bm{\tau}\big)-\big(\vec{\bm{B}}\circ\bm{\tau}\big)\times\frac{d\vec{x}}{dt}\right]+\frac{\tau}{m_{\star}}\left[g_{4}\left(\nabla\bm{m}+2g\,\vec{\bm{\Omega}}\bm{\times}\bm{m}\right)-g_{0}\left(\nabla\bm{\Omega}_{t}+2g\,\vec{\bm{\Omega}}\bm{\times}\bm{\Omega}_{t}\right)\right]\circ\bm{\tau}\\
&\hspace{1cm}+\frac{2\tau}{m_{\star}}\left[g_{2}\bm{m}\circ\nabla\bm{m}+g_{3}\bm{\Omega}_{\mu}\circ\nabla\bm{\Omega}_{\mu}\right]\tau_{0}.\nonumber
\end{align}
To simplify the forthcoming calculations, we recast Eq.~\eqref{EoM_Drude2b} as 
\begin{equation}
\label{EoM_Drude3}
\frac{d\vec{x}_{1}}{dt}=\mu\left[\frac{d\vec{x}_{0}}{dt}\times(\vec{\bm{B}}'\circ\bm{\tau})-(\vec{\bm{B}}'\circ\bm{\tau})\times\frac{d\vec{x}_{0}}{dt}\right]+\mu\vec{C}\tau_{0}+\mu\vec{\bm{A}}\circ\bm{\tau},
\end{equation}
in terms of auxiliary vectors $\vec{\bm{A}},\vec{\bm{B}}'$ and $\vec{C}$. We mention that, in the above expression, we have substituted the time derivative of the position operator in the cross product by $\frac{d\vec{x}_{0}}{dt}$ to maintain the perturbative (first) order of the equation. The expectation value of the charge current then reads
\begin{align}
\label{exp_current0}
\langle\vec{j}_{0}\rangle&=e\textrm{Tr}\left[\hat{\rho}\frac{d\vec{x}_{0}}{dt}\right]=ne\frac{d\vec{x}_{0}}{dt}=\sigma_{D}\vec{E}_{\textrm{ext}}=\vec{j}^{e},\\
\label{exp_currentC}
\langle\vec{j}_{C}\rangle&=e\mu\textrm{Tr}\big[\hat{\rho}\vec{C}\big]=n e\mu \vec{C}=\sigma_{D}\vec{C},\\
\label{exp_currentA}
\langle\vec{j}_{A}\rangle&=e\mu\textrm{Tr}\big[\hat{\rho}(\vec{\bm{A}}\circ\bm{\tau})\big]=\frac{n e\mu}{2}\underbrace{\textrm{Tr}\big[A^{b}_{\alpha}\tau_{b}\big]}_{=0}\hat{e}_{\alpha}+\frac{n e\mu}{2}\textrm{Tr}\big[p^{b}A^{c}_{\alpha}\tau_{b}\tau_{c}\big]\hat{e}_{\alpha}=\frac{ne\mu}{2}2p^{b}A^{b}_{\alpha}\hat{e}_{\alpha}=\sigma_{D}(\bm{p}\circ\vec{\bm{A}}),\\
\label{exp_currentB}
\langle\vec{j}_{B'}\rangle&=2e\mu\textrm{Tr}\left[\hat{\rho}\left\{\textstyle{\frac{d\vec{x}_{0}}{dt}}\times(\vec{\bm{B}'}\circ\bm{\tau})\right\}\right]=2e\mu\epsilon_{\alpha\beta\gamma}\hat{e}_{\alpha}\textstyle{\frac{dx_{0,\beta}}{dt}}B'^{b}_{\gamma}\textstyle{\frac{n}{2}}\textrm{Tr}\big[(\bm{p}\circ\bm{\tau})\tau_{b}\big]=2\mu\vec{j}^{e}\times(\bm{p}\circ\vec{\bm{B}}').
\end{align}
By collecting all the contributions to the current, we obtain the expression
\begin{align}
\label{exp_current}
\vec{j}&=\vec{j}^{e}+\sigma_{D}\big[\vec{C}+\bm{p}\circ\vec{\bm{A}}\,\big]+2\mu\vec{j}^{e}\times(\bm{p}\circ\vec{\bm{B}}'),\\
&=\vec{j}^{e}-\textstyle{\frac{2\sigma_{D}}{|e|}}\left[g_{2}\bm{m}\circ\nabla\bm{m}+g_{3}\bm{\Omega}_{\mu}\circ\nabla\bm{\Omega}_{\mu}\right]+\textstyle{\frac{\sigma_{D}}{|e|}}\bm{p}\circ\left[g_{0}\left(\nabla\bm{\Omega}_{t}+2g\,\vec{\bm{\Omega}}\bm{\times}\bm{\Omega}_{t}\right)-g_{4}\left(\nabla\bm{m}+2g\,\vec{\bm{\Omega}}\bm{\times}\bm{m}\right)\right]\nonumber\\
&\hspace{0.5cm}+\vec{j}^{e}\times\big(\bm{p}\circ\textstyle{\frac{\hbar g\tau}{m_{\star}}}\vec{\bm{B}}\big).\nonumber
\end{align}
The latter term appears to be the topological Hall current in the context of frustrated magnetism. The second and third terms in the above expression can be further simplified by considering the identities~(4) of the main text:
\begin{equation}
\label{rot_frame}
\partial_{t}|_{\textrm{rot}}\vec{\bm{\Omega}}\equiv(\partial_{t}-\bm{\Omega}_{t}\bm{\times})\vec{\bm{\Omega}}=\nabla\bm{\Omega}_{t}+\vec{\bm{\mathcal{J}}},
\end{equation}
where we have used the relation $\partial_{t}\bm{G}|_{\textrm{lab}}=\partial_{t}\bm{G}|_{\textrm{rot}}+\bm{\Omega}_{t}\bm{\times}\bm{G}$ for any vector $\bm{G}$ in spin space. Here, the subindex 'lab' refers to the laboratory frame and 'rot' to the rotating frame at the frequency $\bm{\Omega}_{t}$. Consequently, in the rotating frame of reference, the time derivative of the spatial Yang-Mills fields equals the sum of the gradient of the time Yang-Mills field and the disclination currents. In combination with Eqs.~(7) and~(8) of the main text we obtain
\begin{align}
\label{exp_recast_1}
\nabla\bm{\Omega}_{t}+2g\,\vec{\bm{\Omega}}\bm{\times}\bm{\Omega}_{t}=\partial_{t}\vec{\bm{\Omega}}+\left(1+2g\right)\vec{\bm{\Omega}}\bm{\times}\bm{\Omega}_{t}-\vec{\bm{\mathcal{J}}}=\partial_{t}\vec{\bm{\Omega}}+\vec{\bm{E}}.
\end{align}
On the other hand, the constitutive relations~(3) and~(4) for the disclination density and currents allow us to cast $\bm{\Omega}_{\mu}\circ\partial_{i}\bm{\Omega}_{\mu}=\bm{\Omega}_{\mu}\circ\partial_{\mu}\bm{\Omega}_{i}+\bm{\Omega}_{\mu}\circ\bm{\rho}_{i\mu}$, $\mu=x,y,z,$ and $\bm{\Omega}_{t}\circ\partial_{i}\bm{\Omega}_{t}=\bm{\Omega}_{t}\circ\partial_{t}\bm{\Omega}_{i}-\bm{\Omega}_{t}\circ\bm{\mathcal{J}}_{\hspace{-0.05cm}i}$. Furthermore, we can relate the macroscopic spin density to the time-index Yang-Mills field via the constitutive relation $\bm{m}=\chi\bm{\Omega}_{t}$ of frustrated magnets [51]. As a result, the second and third terms of Eq.~\eqref{exp_current} read
\begin{align}
\label{2nd_term}
&\tfrac{2g_{2}}{|e|}\bm{m}\circ\nabla\bm{m}+\tfrac{2g_{3}}{|e|}\bm{\Omega}_{\mu}\circ\nabla\bm{\Omega}_{\mu}=\tfrac{2}{|e|}\left[g_{2}\chi^{2}\bm{\Omega}_{t}\circ\partial_{t}\vec{\bm{\Omega}}+g_{3}(\vec{\bm{\Omega}}\cdot\nabla)\circ\vec{\bm{\Omega}}\right]+\tfrac{2}{|e|}\left[-g_{2}\chi^{2}\bm{\Omega}_{t}\circ\vec{\bm{\mathcal{J}}}+g_{3}\,\vec{\bm{\Omega}}\circledast\vec{\bm{\varrho}}\right],\\
\label{3rd_term}
&\tfrac{g_{0}}{|e|}\left\{\nabla\bm{\Omega}_{t}+\tfrac{4m_{\star}g_{1}}{\hbar^{2}}(\vec{\bm{\Omega}}\bm{\times}\bm{\Omega}_{t})\right\}-\tfrac{g_{4}}{|e|}\left\{\nabla\bm{m}+\tfrac{4m_{\star}g_{1}}{\hbar^{2}}(\vec{\bm{\Omega}}\bm{\times}\bm{m})\right\}=\tfrac{g_{0}-g_{4}\chi}{|e|}\left[\partial_{t}\vec{\bm{\Omega}}+\vec{\bm{E}}\right].
\end{align}

\section{Topological Hall effect due to SO(3) solitons}

Unit-norm quaternions, $\mathbf{q}=(w,\bm{v})$, offer a convenient parametrization of rotation matrices [51], namely
\begin{align}
\label{eq:Rq}
R_{\alpha\beta}&=\big(1-2\left|\bm{v}\right|^2\big)\delta_{\alpha\beta}+2\,v_{\alpha}v_{\beta}-2\,\varepsilon_{\alpha\beta\gamma}\,w\,v_{\gamma},
\end{align}
where $w$ parametrizes the rotation angle and $\bm{v}$ lies along the rotation axis. This set of quaternions is complemented with the Hamilton product $\mathbf{q}_{1}\wedge\mathbf{q}_{2}\equiv(w_{1}w_{2}-\bm{v}_{1}\circ\bm{v}_{2},w_{1}\bm{v}_{2}+w_{2}\bm{v}_{1}+\bm{v}_{1}\bm{\times}\bm{v}_{2})$ and the (adjoint) inverse operation $\mathbf{q}^{\star}\equiv(w,-\bm{v})$.

\subsection{Topological Hall effect induced by a Shankar skyrmion}

The quaternion parametrization of this SO(3) texture, whose center we locate at the origin of the coordinate system without loss of generality, is given by
\begin{equation}
\label{quater_Shankar_sky}
w=\cos[f(r)/2],\hspace{0.5cm}\bm{v}=\sin[f(r)/2]\hat{\bm{e}}_{r},
\end{equation}
where $f(r)$ is a smooth radial function satisfying the boundary conditions $f(0)=2\pi$ and $f(\infty)=0$. The rigid hard cutoff ansatz for skyrmions, which we will utilize hereafter, is defined by the functional dependence $f(r)=2\pi(1-\tfrac{r}{R_{\star}})\Theta(R_{\star}-r)$ of the rotation angle, where $R_{\star}$ is the radius of the Shankar skyrmion and $\Theta(x)$ is the Heaveside theta function. As it has been discussed in Ref.~50, the following expression 
\begin{align}
\label{Farad_tensor_quater}
\bm{\Omega}_{\mu}\bm{\times}\bm{\Omega}_{\nu}&=2\left(\partial_{\nu}\mathbf{q}\wedge\partial_{\mu}\mathbf{q}^{\star}-\partial_{\mu}\mathbf{q}\wedge\partial_{\nu}\mathbf{q}^{\star}\right)=4\partial_{\mu}w\partial_{\nu}\bm{v}-4\partial_{\nu}w\partial_{\mu}\bm{v}+4\partial_{\mu}\bm{v}\bm{\times}\partial_{\nu}\bm{v},
\end{align}
holds in the absence of topological singularities. With account of the mathematical identities
\begin{align}
\label{inter_math_ident}
&\partial_{\mu}r=\frac{x_{\mu}}{r},\hspace{0.5cm}\partial_{\mu}\hat{\bm{e}}_{r}=\partial_{\mu}\left[\frac{\bm{r}}{r}\right]=\frac{\hat{\bm{e}}_{\mu}}{r}-\frac{x_{\mu}}{r^{2}}\hat{\bm{e}}_{r},\\
&\partial_{\mu}w=-\sin\left[f(r)/2\right]\tfrac{f'(r)}{2}\partial_{\mu}r,\hspace{0.5cm}\partial_{\mu}\bm{v}=\cos\left[f(r)/2\right]\tfrac{f'(r)}{2}(\partial_{\mu}r)\hat{\bm{e}}_{r}+\sin\left[f(r)/2\right]\partial_{\mu}\hat{\bm{e}}_{r},\nonumber
\end{align}
the above equation becomes:
\begin{align}
\label{Farad_tensor_quater2}
\bm{\Omega}_{\mu}\bm{\times}\bm{\Omega}_{\nu}&=4\left[\left(\frac{f'(r)}{4r^{3}}\sin[f(r)]-\frac{1}{r^{4}}\sin^{2}\left[f(r)/2\right]\right)(x_{\alpha}x_{\mu}\epsilon_{\alpha\nu\rho}-x_{\alpha}x_{\nu}\epsilon_{\alpha\mu\rho})+\frac{1}{r^{2}}\sin^{2}[f(r)/2]\epsilon_{\mu\nu\rho}\right]\hat{\bm{e}}_{\rho}\\
&\hspace{0.5cm}+\frac{2f'(r)}{r^{2}}\sin^{2}\left[f(r)/2\right]\left(x_{\nu}\hat{\bm{e}}_{\mu}-x_{\mu}\hat{\bm{e}}_{\nu}\right).\nonumber
\end{align}
The effective magnetic field $\vec{\bm{B}}$, defined by $\bm{B}_{k}=\tfrac{1}{2}\epsilon_{k\mu\nu}\bm{f}_{\mu\nu}$, therefore reads
\begin{equation}
\label{eff_mag_field_spin}
\bm{B}_{k}\propto\sin\left[f(r)\right]\frac{f'(r)}{r}\left(\hat{\bm{e}}_{k}-\frac{x_{k}}{r}\hat{\bm{e}}_{r}\right)+\frac{4x_{k}}{r^{3}}\sin^{2}\left[f(r)/2\right]\hat{\bm{e}}_{r}+\sin^{2}\left[f(r)/2\right]\frac{f'(r)}{r^{2}}\epsilon_{k\mu\nu}(x_{\nu}\hat{\bm{e}}_{\mu}-x_{\mu}\hat{\bm{e}}_{\nu}).
\end{equation}
By projecting $\vec{\bm{B}}$ onto $\hat{\bm{e}}_{x,y,z}$ we obtain the following spatial dependence for the effective magnetic field polarized along the axes of the spin frame of reference:
\begin{align}
\label{eff_mag_field_real}
\vec{B}^{x}&\propto\left(\tfrac{\pi\sin[f]}{2R_{\star}r}\left(\tfrac{x^{2}}{r^{2}}-1\right)+\tfrac{x^{2}}{r^{4}}\sin^{2}\big[\tfrac{f}{2}\big], \tfrac{\pi z }{R_{\star}r^{2}}\sin^{2}\big[\tfrac{f}{2}\big]+\tfrac{\pi xy}{2R_{\star}r^{3}}\sin\left[f\right]+\tfrac{xy}{r^{4}}\sin^{2}\big[\tfrac{f}{2}\big],-\tfrac{\pi y}{R_{\star}r^{2}}\sin^{2}\big[\tfrac{f}{2}\big]+\tfrac{\pi xz}{2R_{\star}r^{3}}\sin\left[f\right]+\tfrac{xz}{r^{4}}\sin^{2}\big[\tfrac{f}{2}\big]\right),\\
\vec{B}^{y}&\propto\left(-\tfrac{\pi z}{R_{\star}r^{2}}\sin^{2}\big[\tfrac{f}{2}\big]+\tfrac{\pi xy}{2R_{\star}r^{3}}\sin\left[f\right]+\tfrac{xy}{r^{4}}\sin^{2}\big[\tfrac{f}{2}\big], \tfrac{\pi\sin[f]}{2R_{\star}r}\left(\tfrac{y^{2}}{r^{2}}-1\right)+\tfrac{y^{2}}{r^{4}}\sin^{2}\big[\tfrac{f}{2}\big],\tfrac{\pi x}{R_{\star}r^{2}}\sin^{2}\big[\tfrac{f}{2}\big]+\tfrac{\pi yz}{2R_{\star}r^{3}}\sin\left[f\right]+\tfrac{yz}{r^{4}}\sin^{2}\big[\tfrac{f}{2}\big]\right),\nonumber\\
\vec{B}^{z}&\propto\left(\tfrac{\pi y}{R_{\star}r^{2}}\sin^{2}\big[\tfrac{f}{2}\big]+\tfrac{\pi xz}{2R_{\star}r^{3}}\sin\left[f\right]+\tfrac{xz}{r^{4}}\sin^{2}\big[\tfrac{f}{2}\big], -\tfrac{\pi x}{R_{\star}r^{2}}\sin^{2}\big[\tfrac{f}{2}\big]+\tfrac{\pi yz}{2R_{\star}r^{3}}\sin\left[f\right]+\tfrac{yz}{r^{4}}\sin^{2}\big[\tfrac{f}{2}\big],\tfrac{\pi\sin[f]}{2R_{\star}r}\left(\tfrac{z^{2}}{r^{2}}-1\right)+\tfrac{z^{2}}{r^{4}}\sin^{2}\big[\tfrac{f}{2}\big]\right).\nonumber
\end{align} 
where we have accounted for the hard cutoff ansatz introduced above and rescaled the magnetic field by a factor of 4. 
The expressions for the averaged (polarized) components of the magnetic field are, after some algebra, $\vec{B}^{x}=\vec{B}^{y}=\vec{B}^{z}=\vec{0}$, so that Shankar skyrmions do not mediate a topological Hall response in frustrated magnets.

\subsection{Topological Hall effect induced by a(n Anderson-Toulouse) $4\pi$-vortex}

The quaternion parametrization of this SO(3) texture, whose axis lies along the $z$-direction, is given by
\begin{equation}
\label{quater_AT_vortex}
w=\cos[f(\rho)/2],\hspace{0.5cm}\bm{v}=\sin[f(\rho)/2]\hat{\bm{e}}_{\phi},
\end{equation}
where $f(\rho)$ is a smooth radial function satisfying the boundary conditions $f(0)=0$ and $f(R_{\star})=\pi$, and $\hat{\bm{e}}_{\phi}=(-\sin\phi,\cos\phi,0)$. Here, $R_{\star}$ is the radius of the $4\pi$-vortex and $\phi$ denotes the azimuthal angle in cylindrical coordinates. The rigid hard cutoff ansatz for $4\pi$-vortices that we will utilize hereafter is defined by the functional dependence $f(\rho)=\tfrac{\pi}{R_{\star}}\rho\,\Theta(R_{\star}-\rho)+\pi\Theta(\rho-R_{\star})$ of the rotation angle, where $\rho=\sqrt{x^{2}+y^{2}}$ is the radius in cylindrical coordinates and $\Theta(x)$ is the Heaveside theta function. With account of the mathematical identities
\begin{align}
\label{inter_math_ident2}
&\partial_{\mu}\rho=\frac{x_{\mu}}{\rho},\hspace{0.5cm}\partial_{\mu}\hat{\bm{e}}_{\phi}=-\left(\partial_{\mu}\phi\right)\hat{\bm{e}}_{\rho},\hspace{0.5cm}\hat{\bm{e}}_{\rho}=(\cos\phi,\sin\phi,0),\\
&\partial_{\mu}w=(\delta_{\mu,z}-1)\sin\left[f(\rho)/2\right]\tfrac{f'(\rho)x_{\mu}}{2\rho},\hspace{0.5cm}\partial_{\mu}\bm{v}=(1-\delta_{\mu,z})\left\{\cos\left[f(\rho)/2\right]\tfrac{f'(\rho)x_{\mu}}{2\rho}\hat{\bm{e}}_{\phi}+\sin\left[f(\rho)/2\right]\partial_{\mu}\hat{\bm{e}}_{\phi}\right\},\nonumber
\end{align}
Eq.~\eqref{Farad_tensor_quater} becomes:
\begin{align}
\label{Farad_tensor_quater3}
\bm{\Omega}_{\mu}\bm{\times}\bm{\Omega}_{\nu}&=(1-\delta_{\mu,z})(1-\delta_{\nu,z})\left(2\sin^{2}\left[f(\rho)/2\right]\hat{\bm{e}}_{\rho}+\sin\left[f(\rho)\right]\hat{\bm{e}}_{z}\right)\frac{f'}{\rho^{3}}(\epsilon_{\nu z\lambda}x_{\mu}-\epsilon_{\mu z\lambda}x_{\nu})x_{\lambda}.
\end{align}
By recalling again the definition $\bm{B}_{k}=\tfrac{1}{2}\epsilon_{k\mu\nu}\bm{f}_{\mu\nu}$ for the spatial components of the effective magnetic field, we obtain
\begin{align}
\label{eff_mag_field_spin2}
\bm{B}_{x}&=\bm{B}_{y}=\bm{0},\hspace{0.5cm}\bm{B}_{z}\propto\frac{f'}{\rho}\left(2\sin^{2}\left[f(\rho)/2\right]\hat{\bm{e}}_{\rho}+\sin\left[f(\rho)\right]\hat{\bm{e}}_{z}\right).
\end{align} 
The projection of $\vec{\bm{B}}$ onto $\hat{\bm{e}}_{x,y,z}$, yields the following spatial dependence for the components of the effective magnetic field polarized along the axes of the spin frame of reference:
\begin{align}
\label{eff_mag_field_real}
\vec{B}^{x}&\propto\left(0,0,\frac{2f'x}{\rho^{2}}\sin^{2}\left[f(\rho)/2\right]\right),\\
\vec{B}^{y}&\propto\left(0,0,\frac{2f'y}{\rho^{2}}\sin^{2}\left[f(\rho)/2\right]\right),\nonumber\\
\vec{B}^{z}&\propto\left(0,0,\frac{f'}{\rho}\sin\left[f(\rho)\right]\right),\nonumber
\end{align}
and, with account of the hard cutoff ansatz introduced above, we finally obtain the expression 
\begin{equation}
\label{eff_mag_field_averaged}
\big\langle\vec{B}^{x}\big\rangle=\big\langle\vec{B}^{y}\big\rangle=\vec{0},\hspace{0.5cm}\big\langle\vec{B}^{z}\big\rangle=4\left(1+2g\right)\frac{\hat{e}_{z}}{R_{\star}^{2}},
\end{equation} 
for the volume-averaged spin-polarized components of the effective magnetic field.

\section{Magnetic disclination density}

The Yang-Mills fields are cast as $\bm{\Omega}_{\mu}=2\partial_{\mu}\mathbf{q}\wedge\mathbf{q}^{\star}$ in the quaternion representation. As a result, a direct calculation yields the following expression for the components of the magnetic disclination tensor:
\begin{equation}
\label{mag_discl_tensor}
\bm{\rho}_{\mu\nu}=2(\partial_{\mu}\partial_{\nu}\mathbf{q}-\partial_{\nu}\partial_{\mu}\mathbf{q})\wedge\mathbf{q}^{\star}=2w(\partial_{\mu\nu}\bm{v}-\partial_{\nu\mu}\bm{v})-(\partial_{\mu\nu}w-\partial_{\nu\mu}w)\bm{v}+2\bm{v}\bm{\times}(\partial_{\mu\nu}\bm{v}-\partial_{\nu\mu}\bm{v}).
\end{equation}
In the main text we have considered a magnetic disclination line extending along the $z$ axis; we take hereafter $\vec{r}_{d}=(x_{d},y_{d},z)$ as its parametrization. By following Ref.~62, we describe this topological defect via the quaternion $\mathbf{q}_{d}=\big(\cos(s\phi/2),\sin(s\phi/2)\hat{e}_{z}\big)$, where $\phi=\tan^{-1}[y-y_{d}/x-x_{d}]+\phi_{0}$ is the azimuthal angle and $s=\Omega/2\pi$ denotes the strength of the disclination line. By incorporating this parametrization into Eq.~\eqref{mag_discl_tensor}, we derive
\begin{equation}
\label{mag_discl_tensor2}
\bm{\rho}_{xy}=s(\partial_{x}\partial_{y}\phi-\partial_{y}\partial_{x}\phi)\bm{\hat{e}}_{z},\hspace{0.5cm}\bm{\rho}_{xz}=\bm{\rho}_{yz}=\bm{0}.
\end{equation}
Since $\partial_{x}\partial_{y}\phi-\partial_{y}\partial_{x}\phi=2\pi s\delta(\vec{r}-\vec{r}_{d})$, and with account of the definition of the disclination density vector in terms of the corresponding tensor, we obtain the expression appearing in the main text.


\begin{thebibliography}{99}

\bibitem{Kibble-JPA1976} T.W.B. Kibble, J Phys. A: Math. Gen. {\bf 9}, 1387 (1976).
\bibitem{Kibble-PR1980} T.W.B. Kibble, Phys. Rep. {\bf 67}, 183 (1980).
\bibitem{Zurek-Nature1985} W.H. Zurek, Nature {\bf 317}, 505 (1985).
\bibitem{Zurek-APPB1993} W.H. Zurek, Acta Phys. Pol. B {\bf 24}, 1301 (1993).
\bibitem{Zurek-PR1996} W.H. Zurek, Phys Rep. {\bf 276}, 177 (1996).
\bibitem{delCampo-IJMPA2014} A. del Campo and W.H. Zurek, Int. J. Mod. Phys. A {\bf 29}, 1430018 (2014).
\bibitem{Romanov-PSSA1983} A.E. Romanov and V.I. Vladimirov, Phys. Stat. Sol. (a) {\bf 78}, 11 (1983).
\bibitem{Kleman-RMP2008} M. Kleman and J. Friedel, Rev. Mod. Phys. {\bf 80}, 61 (2008). 
\bibitem{Lavrentovich-Springer2001} O. Lavrentovich, P. Pasini, C. Zannoni, and S. Zumer (Eds.), \textit{Defects in Liquid Crystals: Computer Simulations, Theory and Experiments}, Springer, Boston (2001).
\bibitem{Andrienko-JML2018} D. Andrienko, J. Mol. Liq. {\bf 267}, 520 (2018).
\bibitem{Toulouse-CommPhys1977} G. Toulouse, Commun. Phys. {\bf 2}, 115 (1977).
\bibitem{Villain-JPhys1977} J. Villain, J. Phys. C {\bf 10}, 1717 (1977).
\bibitem{Toulouse-PhysRep1979} G. Toulouse, Phys. Rep. {\bf 49}, 267 (1979).
\bibitem{Henley-AnnPhys1984a} C.L. Henley, Ann. Phys. {\bf 156}, 368 (1984).
\bibitem{Edwards-JPF1975} S.F. Edwards and P.W. Anderson, J. Phys. F: Metal Phys. {\bf 5}, 965 (1975).
\bibitem{Shankar-JPhys1977} R. Shankar, Journal de Physique {\bf 38}, 1405 (1977).
\bibitem{Volovik-JETP1977} G.E. Volovik and V.P. Mineev, Sov. Phys. JETP {\bf 46}, 401 (1977).
\bibitem{Anderson-PRL1976} P.W. Anderson and G. Toulouse, Phys. Rev. Lett. {\bf 38}, 508 (1976).
\bibitem{Mermin-RMP1979} N.D. Mermin, Rev. Mod. Phys. {\bf 51}, 591 (1979).
\bibitem{Zarzuela-PRB2021} R. Zarzuela and J. Sinova, arXiv:2107.13330 (2021).
\bibitem{Ralph-JMMM2008} D.C. Ralph and M.D. Stiles, J. Magn. Magn. Mater. {\bf 320}, 1190 (2008).
\bibitem{Gomonay-LTP2014} E.V. Gomonay and V.M. Loktev, Low Temp. Phys. {\bf 40}, 17 (2014).
\bibitem{Baltz-RMP2018} V. Baltz, A. Manchon, M. Tsoi, T. Moriyama, T. Ono, and Y. Tserkovnyak, Rev. Mod. Phys. {\bf 90}, 015005 (2018).
\bibitem{Taguchi-Science2001} Y. Taguchi, Y. Oohara, H. Yoshizawa, N. Nagaosa, and Y. Tokura, Science {\bf 291}, 2573 (2001).
\bibitem{Yanagihara-PRL2002} H. Yanagihara and M.B. Salamon, Phys. Rev. Lett. {\bf 89}, 187201 (2002).
\bibitem{Machida-PRL2007} Y. Machida, S. Nakatsuji, Y. Maeno, T. Tayama, T. Sakakibara, and S. Onoda, Phys. Rev. Lett. {\bf 98}, 057203 (2007).
\bibitem{Neubauer-PRL2009} A. Neubauer, C. Pfleiderer, B. Binz, A. Rosch, R. Ritz, P.G. Niklowitz, and P. B\"{o}ni, Phys. Rev. Lett. {\bf 102}, 186602 (2009).
\bibitem{Kanazawa-PRL2011} N. Kanazawa, Y. Onose, T. Arima, D. Okuyama, K. Ohoyama, S. Wakimoto, K. Kakurai, S. Ishiwata, and Y. Tokura, Phys. Rev. Lett. {\bf 106}, 156603 (2011).
\bibitem{Yu-NatMater2011} X.Z. Yu, N. Kanazawa, Y. Onose, K. Kimoto, W.Z. Zhang, S. Ishiwata, Y. Matsui, and Y. Tokura, Nat. Mater. {\bf 10}, 106 (2011).
\bibitem{Ueda-PRL2012} K. Ueda, S. Iguchi, T. Suzuki, S. Ishiwata, Y. Taguchi, and Y. Tokura, Phys. Rev. Lett. {\bf 108}, 156601 (2012).
\bibitem{Huang-PRL2012} S.X. Huang and C.L. Chien, Phys. Rev. Lett. {\bf 108}, 267201 (2012).
\bibitem{Shiomi-PRB2012} Y. Shiomi, S. Iguchi, and Y. Tokura, Phys. Rev. B {\bf 86}, 180404(R) (2012). 
\bibitem{Li-PRL2013} Y. Li, N. Kanazawa, X.Z. Yu, A. Tsukazaki, M. Kawasaki, M. Ichikawa, X.F. Jin, F. Kagawa, and Y. Tokura, Phys. Rev. Lett. {\bf 110}, 117202 (2013).
\bibitem{Surgers-NatComms2014} C. S\"{u}rgers, G. Fisher, P. Winkel, and H. v. L\"{o}hneysen, Nat. Comms. {\bf 5}, 3400 (2014).
\bibitem{Leroux-SciRep2018} M. Leroux, M.J. Stolt, S. Jin, D.V. Pete, C. Reichhardt, and B. Maiorov, Sci. Rep. {\bf 8}, 15510 (2018).
\bibitem{Kurumaji-Science2019} T. Kurumaji, T. Nakajima, M. Hirschberger, A. Kikkawa, Y. Ymasaki, H. Sagayama, H. Nakao, Y. Taguchi, T. Arima, and Y. Tokura, Science {\bf 365}, 914 (2019). 
\bibitem{Shao-NatElectron2019} Q. Shao, Y. Liu, G. Yu, S.K. Kim, X. Che, C. Tang, Q.L. He, Y. Tserkovnyak, J. Shi, and K.L. Wang, Nat. Electron. {\bf 2}, 182 (2019).
\bibitem{Yu-JMMM2019} J. Yu, L. Liu, J. Deng, C. Zhou, H. Liu, F. Poh, and J. Chen, J. Mag. Mag. Mat. {\bf 487}, 165316 (2019).
\bibitem{Zhang-APL2019} S. Zhang, S. Xia, Q. Cao, D. Wang, R. Liu, and Y. Du, Appl. Phys. Lett. {\bf 115}, 022404 (2019). 
\bibitem{Li-APL2019} H. Li, B. Ding, J. Chen, Z. Li, Z. Hou, E. Liu, H. Zhang, X. Xi, G. Wu, and W. Wang, Appl. Phys. Lett. {\bf 114}, 192408 (2019). 
\bibitem{He-APL2020} Y. He, J. Kroder, J. Gayles, C. Fu, Y. Pan, W. Schnelle, C. Felser, and G.H. Fecher, Appl. Phys. Lett. {\bf 117}, 052409 (2020).
\bibitem{Lim-Small2020} Z.S. Lim, C. Li, Z. Huang, X. Chi, J. Zhou, S. Zeng, G.J. Omar, Y.P. Feng, A. Rusydi, S.J. Pennycook, T. Venkatesan, and A. Ariando, Small {\bf 16}, 2004683 (2020).
\bibitem{Ghimire-SciAdv2020} N.J. Ghimire, R.L. Dally, L. Poudel, D.C. Jones, D. Michel, N.Thapa Magar, M. Bleuel, M.A. McGuire, J.S. Jiang, J.F. Mitchell, J.W. Lynn, and I.I. Mazin, Sci. Adv. {\bf 6} (51), eabe2680 (2020). 
\bibitem{Wang-PRB2021} Q. Wang, K.J. Neubauer, C. Duan, Q. Yin, S. Fujitsu, H. Hosono, F. Ye, R. Zhang, S. Chi, K. Krycka, H. Lei, and P. Dai, Phys. Rev. B {\bf 103}, 014416 (2021). 
\bibitem{Dzyaloshinskii-AnnPhys1980} I.E. Dzyaloshinskii and G.E. Volovick, Ann. Phys. {\bf 125}, 67 (1980).
\bibitem{Volovik-ZhETF1978} G.E. Volovik and I.E. Dzyaloshinskii, Zh. Eksp. Teor. Fiz. {\bf 75}, 1102 (1978).
\bibitem{Footnote00} In general, an Euclidean medium exhibiting geometric frustration (e.g, liquid crystals, amorphous metals, etc.) can be mapped onto a Riemannian manifold with curvature engendered by the disclination lines/loops \cite{Kondo-1955,Bilby-1960}.
\bibitem{Kondo-1955} K. Kondo, \textit{RAAG Memoirs of the Unifying Study of Basic Problems in Engineering and Physical Sciences by Means of Geometry}, Gakujutsu Bunken Fukyu-Kai, Tokyo (1955-1967).
\bibitem{Bilby-1960} B.A. Bilby, Prog. Solid Mech. {\bf 1}, 329 (1960).
\bibitem{Zarzuela-PRB2019} R. Zarzuela, H. Ochoa and Y. Tserkovnyak, Phys. Rev. B {\bf 100}, 054426 (2019).
\bibitem{Ochoa-PRB2018} H. Ochoa, R. Zarzuela, and Y. Tserkovnyak, Phys. Rev. B {\bf 98}, 054424 (2018).
\bibitem{Yang-PR1954} C. N. Yang and R. L. Mills, Phys. Rev. {\bf 96}, 191 (1954).
\bibitem{FN1} The general expression for the components of the Faraday tensor reads $f_{\mu\nu}^{a}=\partial_{\mu}\Omega_{\nu}^{a}-\partial_{\nu}\Omega_{\mu}^{a}+gf^{abc}\Omega_{\mu}^{b}\Omega_{\nu}^{c}$, where $\Omega_{\mu}^{a}$ denotes the vector potential and $f^{abc}$ are the structure constants of the Lie algebra associated with the nonabelian Lie group of the Yang-Mills theory. The later defines the commutation relations $[T^{a},T^{b}]=if^{abc}T^{c}$ among the generators $T^{a}$ of the Lie algebra. Our expression is derived by accounting for the commutator algebra $[\tau^{a},\tau^{b}]=2i\epsilon_{abc}\tau^{c}$ for Pauli matrices.
\bibitem{SM} See Supplemental Material.
\bibitem{Chudnovsky-PRL2007} E. M. Chudnovsky, Phys. Rev. Lett. {\bf 99}, 206601 (2007).
\bibitem{FN2} Smoothness of the O(3) walls implies that the determinant of the order parameter vanishes identically across the wall center (which, in turn, implies the vanishing of one of the singular values of the rotation matrix, see Ref.~\onlinecite{Henley-AnnPhys1984a} for further details).
\bibitem{Wornle-2019} M.S. W\"{o}rnle, P. Welter, Z. Ka\v{s}par, K. Olejn\'{i}k, V. Nov\'{a}k, R.P. Campion, P. Wadley, T. Jungwirth, C.L. Degen, P. Gambardella, arXiv:1912.05287 (2019).
\bibitem{FN5} In this section we focus on \textit{wedge} magnetic disclinations of angle $\Omega$. In principle, the rotation angle could take any value in the range $|\Omega|\leq\pi$, but some numerical simulations of topological defects in frustrated magnets point towards the stabilization of disclination lines of strength $\Omega/2\pi=\pm1$, see Ref. \onlinecite{Henley-AnnPhys1984b}.
\bibitem{Henley-AnnPhys1984b} C.L. Henley, Ann. Phys. {\bf 156}, 324 (1984).
\bibitem{Callan-Jones-PRB2006} A.C. Callan-Jones, R.A. Pelcovits, V.A. Slavin, S. Zhang, D.H. Laidlaw, and G.B. Loriot, Phys. Rev. E {\bf 74}, 061701 (2006).
\bibitem{Jin-JPA2006} P.-Q. Jin, Y.-Q. Li and F.-C. Zhang, J. Phys. A: Math. Gen. {\bf 39}, 7115 (2006).
\bibitem{Sethna-PRB1985} J.P. Sethna, Phys. Rev. B {\bf 31}, 6278 (1985).
\end{thebibliography}
\end{document}